\documentclass[conference]{IEEEtran}

\usepackage{cite}

\ifCLASSINFOpdf

\else

\fi
\usepackage{amsmath}
\usepackage{cite}
\input{epsf}
\begin{document}

\title{Bounds on the Capacity of ASK Molecular Communication Channels with ISI}

\author{\IEEEauthorblockN{Siavash Ghavami$^{*,**}$, Raviraj Adve$^{**}$, Farshad Lahouti$^{*,***}$,}
\IEEEauthorblockA{$^*$School of Electrical and Computer Engineering, University of Tehran, Tehran, Iran\\
$^{**}$ Department of Electrical and Computer Engineering, University of Toronto, Toronto, Canada\\
$^{***}$ Department of Electrical Engineering, California Institute of Technology, Pasadena, CA\\
Emails: s.ghavami@ut.ac.ir, rsadve@comm.utoronto.ca, lahouti@ut.ac.ir}}
\maketitle

\begin{abstract}
There are now several works on the use of the additive inverse Gaussian noise (AIGN) model for the random transit time in molecular communication~(MC) channels. The randomness invariably causes inter-symbol interference (ISI) in MC, an issue largely ignored or simplified. In this paper we derive an upper bound and two lower bounds for MC based on amplitude shift keying (ASK) in presence of ISI. The Blahut-Arimoto algorithm~(BAA) is modified to find the input distribution of transmitted symbols to maximize the lower bounds. Our results show that over wide parameter values the bounds are close.
\end{abstract}


\IEEEpeerreviewmaketitle
\section{Introduction}
Molecular communications~(MC) systems encode information in the concentration~\cite{1}, time of release~\cite{2}, the number of molecules released in a time-slot and the type and ratio of molecules~\cite{32}. This technique has potential applications in environments where electromagnetic waves cannot be used, e.g., in buried pipelines or in medical applications with embedded devices. This paper considers the case where information is encoded in the number of molecules released in each time-slot, i.e., molecular amplitude shift keying (ASK).

The propagation of molecules in a fluid medium is governed by Brownian motion~\cite{3}, possibly with a drift velocity~\cite{2}. In~\cite{2} such propagation was analyzed and it was shown that the molecules experience a propagation delay that follows an inverse Gaussian distribution. Based on this so-called inverse Gaussian noise (IGN) channel, in \cite{2,4} expressions and bounds for channel capacity are derived when information is encoded on the release time of molecules. The capacity of MC when information is encoded on the concentration of molecules is studied for binary communications in~\cite{5,6} and for binary and 4-ary communications in~\cite{7}. In these works, the aggregate distribution of the number of arrived molecules in a time slot is approximated by a Gaussian distribution. 

In  molecular ASK modulation, molecules diffuse to the receiver where they are detected and removed from the system. One of the main challenges in such a diffusion-based communication system is the resulting inter-symbol interference (ISI); due to the random propagation time, molecules may arrive over many time-slots. Of interest here, therefore, is a capacity analysis of a more likely system which suffers from ISI. In~\cite{3}, disregarding the presence of ISI, a binary ASK scheme by considering the life expectancy of molecule with AIGN model for propagation time is studied (output symbols are independent). Indeed, if the system does not suffer from ISI, in model wherein molecule release and detections are perfect, communications would also be perfect. In~\cite{31}, the capacity and the probability of error of binary and 4-ary ASK schemes are investigated in presence of ISI. However, for capacity analysis, researchers consider maximization of symbol-by-symbol mutual information, which is a lower bound for $I_{i.i.d}$ or the achievable information rate for ISI channel with independent, identically distributed (i.i.d) inputs~\cite{10}.

 Capacity analyses of conventional ISI communications has a strong presence in the literature. Starting from the seminal work in~\cite{8}, several methods have been proposed to derive the capacity of an ISI channel. The problem of determining the channel capacity is completely solved in the case of an unrestricted input distribution but remains open for distributions over a discrete input alphabet~\cite{9}. Several lower and upper bounds to the achievable rate with discrete inputs and Gaussian channels have been obtained in the literature~\cite{10,11}. A simulation-based approach is applicable for calculating capacity in the case of i.i.d.~inputs with binary modulation \cite{14,15}. This approach is based on a trellis whose number of states corresponds to the memory of the ISI channel.

In this work, we consider a general molecular ASK scheme in presence of diffusion-induced ISI. Time is slotted and the information is encoded in the number of molecules released in a time-slot, and the receiver counts the number of molecules received within each time-slot. We obtain the probability of molecules arriving within a specific time-slot using the AIGN model, in turn leading to a binomial distribution on the number of received molecules in each time-slot. We propose two lower bounds and an upper bound for the $I_{i.i.d}$ of such a channel, which is, here, called capacity. To provide a tractable analysis we restrict the effect of ISI to one time-slot, i.e., we assume the molecules that do not arrive within two time-slots have disappeared. The first lower bound is based on a symbol-by-symbol maximization of mutual information, while the second lower bound uses information of next received symbol for decoding. The lower bounds are maximized by optimizing the input distribution, which in turn closes the gap between the lower and upper bounds.

The remainder of this paper is organized as follows, the system under consideration is presented in Section II. The channel characteristics including the effect of ISI are developed in Section III. In Section IV an upper bound and two lower bounds of MC with ISI are proposed including an optimization of these bounds. Section V presents some numerical results; finally Section VI wraps up the paper with some concluding remarks.
\section{System Model}
In our system, the transmitter is a point source of identical molecules. At the beginning of every time slot, with length of $T$, it conveys a message by releasing $X$ molecules into a fluid medium. Here, $X$ with $0 \le X \le {X_{\max }}$, is a random variable and $X_{\max}$ is the maximum number of molecules released in any time slot. The transmitter therefore uses amplitude shift keying (ASK), i.e., the message is encoded in the number of molecules released. Once released, the transmitter does not affect the propagation of the molecules.

Molecules propagate between the transmitter and receiver by Brownian motion characterized by a diffusion constant $d$ and (positive) drift velocity $v$. At the receiver, all received molecules are absorbed and removed from the system. Importantly, for our analysis, we assume that a molecule transmitted in time-slot $m$ either arrives in the same time-slot, the next time-slot $(m+1)$ or disappears. Essentially, for tractability, our analysis focuses on ISI within one time-slot. We assume that everything else within the system operates perfectly, i.e., the only randomness in our model is the propagation time (causing a randomness in the number of molecules that are received). Using the AIGN analysis in~\cite{2}, the cumulative distribution function (CDF) of the propagation time is given by
\begin{eqnarray}\label{eq:1}
{F_W}\left( w \right) & = & \Phi \left( {\sqrt {\frac{\lambda }{w}} \left( {\frac{w}{\mu } - 1} \right)} \right) +
\\ & & \hspace*{0.2in} {e^{\frac{{2\lambda }}{\mu }}}\Phi \left( { - \sqrt {\frac{\lambda }{w}} \left( {\frac{w}{\mu } + 1} \right)} \right), \hfill w > 0. \nonumber
\end{eqnarray}
Here $\Phi \left(  \cdot  \right)$ is the CDF of a standard Gaussian random variable; also, $\mu = l/v$ and $\lambda = l^2/\sigma^2$ in which $l$ is the distance between transmitter and receiver and $\sigma^2 = d/2$ is the variance of the associated Weiner process~\cite{2} and $d$ is the diffusion coefficient.

Let $q_k$ denote the probability that a molecule arrives within the $k$th time-slot following its transmission, i.e., a molecule transmitted at time 0 arrives in time interval $\left((k-1)T,kT\right]$, where $T$ is the time-slot duration. We have ${q_k} = {F_W}\left( {kT} \right) - {F_W}\left( {\left( {k - 1} \right)T} \right)$. The probability of a molecule arriving in the same time slot that it is released is ${q_1} = {F_W}\left( T \right)$.

 The probability of receiving $Y_m = y$ molecules at receiver, when transmitter release $X_m = x$  molecules is binomial:
\begin{equation} \label{eq:2}
\Pr \left( {\left. {{Y_m} = y} \right|{X_m} = x} \right) = \left\{ {\begin{array}{*{20}{c}}
{\left( {\begin{array}{*{20}{c}}
x\\
y
\end{array}} \right)q_1^y{{\left( {1 - {q_1}} \right)}^{x - y}}{\rm{ }},{\rm{0}} \le y \le x}\\
{\begin{array}{*{20}{c}}
0&{}&{}&{}&{}
\end{array}{\rm{,}}y < 0,y > x}
\end{array}} \right.
\end{equation}
We denote by $a_x, 1 \le x \le X_{\max}$, the probability of the transmitter releasing $X_m = x$ molecules. While~\eqref{eq:2} ignores the effect of ISI, we consider its impact in coming sections.

\section{Communications in the Presence of ISI}
With ISI we must consider transmit/receive sequences and the average mutual information per channel use is given by
\begin{equation} \label{eq:3}
I(X^L; Y^{L+k}) = \mathop {\lim }\limits_{L \to \infty } \frac{1}{L}I\left( {{X^L};{Y^{L+k}}} \right),
\end{equation}
where, ${X^L} = \left[ {{X_1},...,{X_L}} \right]$ and ${Y^{L+k}} = \left[ {{Y_1},...,{Y_{L+k}}} \right]$ , denote input and output sequences of length $L$ and $L+k$ evaluated for a given input distribution ${P_{{X^L}}}\left( {{X^L}} \right)$. This determines the achievable rate of reliable communication through this channel with this specific input distribution and the channel capacity is the supremum of this mutual information over all allowed input distributions. This is in contrast to the case of a discrete memoryless channel (DMC) without ISI and all molecules either arrive within the time-slot of transmission or disappear. We denote the maximum value of the mutual information of this DMC as ${C_{DMC}}$.

For tractability, we will focus on the case of independent and i.i.d. inputs, i.e., ${P_{{X^L}}} = {\bf{a}} \times ... \times {\bf{a}} = {{\bf{a}}^L}$, where ${\bf{a}} = \left[ {{a_0},...,{a_{{X_{\max }}}}} \right]$, a simplification suggested by the work in~\cite{10}. We denote the resulting average mutual information as ${I_{i.i.d}}$. Note that this mutual information is a lower bound of the expression in~\eqref{eq:3}, but with a slight abuse of notation we will call the resulting mutual information the ``channel capacity" given by
\begin{equation}\label{eq:4}
C = \mathop {\lim }\limits_{L \to \infty } \frac{1}{L}\mathop {\sup }\limits_{{{\bf{a}}^L}} I\left( {{X^L};{Y^{L+k}}} \right).
\end{equation}

Despite this significant simplification, calculating the channel capacity in \eqref{eq:4} appears intractable and, in this paper, we develop lower and upper bounds on this expression.

At the start of time slot $k$, ${X_k} \in \left\{ {0,1,...,{X_{\max }}} \right\}$  molecules are released. We begin by considering the interference from  $k-1$ previous time-slots. Let ${P_k}\left( n \right)$  denote the probability that $n$  molecules arrive from the $k-1$  previous time-slots, in time slot $k$. Furthermore, let $Z_k$ denote the number of transmitted molecules (at time-slot $k$) \emph{not received} in the same slot and let $N_{k-1}$ denote the total number of interfering molecules from the previous $(k-1)$ time slots, i.e., $N_{k-1} \in \left\{ {0,1,...,{(k - 1) X_{\max }}} \right\}$. Finally, letting $Y_k$ denote the number of molecules received in time-slot $k$ we have
\begin{equation}\label{eq:5}
 Y_k = X_k - Z_k + N_{k-1},
\end{equation}
we have ${Y_k} \in \left\{ {0,1,...,{X_{\max }}k} \right\}$. The transition probability when  ${X_k} = x $ molecules are transmitted, it is given by
\begin{equation}\label{eq:6}
\begin{array}{l}
{p_{\left. {{Y_k}} \right|{X_k}}}\left( {\left. y \right|x} \right) = \\
{\left( {1 - {q_1}} \right)^x}{P_k}\left( y \right) + \left( {\begin{array}{*{20}{c}}
x\\
1
\end{array}} \right){q_1}{\left( {1 - {q_1}} \right)^{x - 1}}{P_k}\left( {y - 1} \right) + ...\\
\left( {\begin{array}{*{20}{c}}
x\\
{x - 1}
\end{array}} \right)q_1^{x - 1}\left( {1 - {q_1}} \right){P_k}\left( {y - \left( {x - 1} \right)} \right) + {q_1}^x{P_k}\left( {y - x} \right),{\rm{   }}\\
y = 0,...,{X_{\max }}k - \left( {{X_{\max }} - x} \right).
\end{array}
\end{equation}
Probability of receiving $y$ molecules for $k-1$  time slots, ${P_k}\left( y \right)$, $y = 0,1,...,{X_{\max }}k - {X_{\max }}$  , simply can be calculated by induction. Hence we have
\begin{equation}\label{eq:7}
\begin{array}{l}
{P_k}\left( y \right) = \left( {{a_0} + ... + {{\left( {1 - {q_k}} \right)}^{{X_{\max }}}}{a_{{X_{\max }}}}} \right){P_{k - 1}}\left( y \right)\\
 + \left( {{q_k}{a_1} + ...{\rm{ + }}\left( {\begin{array}{*{20}{c}}
{{X_{\max }}}\\
1
\end{array}} \right){q_k}{{\left( {1 - {q_k}} \right)}^{{X_{\max }} - 1}}{a_{{X_{\max }}}}} \right) \times \\
{P_{k - 1}}\left( {y - 1} \right) + \\
 \vdots \\
 + \left( {q_k^{{X_{\max }} - 1}{a_{{X_{\max }} - 1}} + \left( {\begin{array}{*{20}{c}}
{{X_{\max }}}\\
{{X_{\max }} - 1}
\end{array}} \right)q_k^{{X_{\max }} - 1}\left( {1 - {q_k}} \right)} \right. \times \\
\left. {{a_{{X_{\max }}}}} \right){P_{k - 1}}\left( {y - \left( {{X_{\max }} - 1} \right)} \right) + \\
q_k^{{X_{\max }}}{a_{{X_{\max }}}}{P_{k - 1}}\left( {y - {X_{\max }}} \right),
\end{array}
\end{equation}
where  ${a_i} = \Pr\left( {{x_1} = i} \right)$  for $i \in \left\{ {0,...,{X_{\max }}} \right\}$. Using induction we can compute ${P_k}\left( y \right)$. In general, we can write
\begin{equation}\label{eq:8}
{P_k}\left( 0 \right) = \left( {{a_0} + ... + {{\left( {1 - {q_k}} \right)}^{{X_{\max }}}}{a_{{X_{\max }}}}} \right){P_{k - 1}}\left( 0 \right),
\end{equation}
\begin{equation}\label{eq:9}
\begin{array}{l}
{P_k}\left( 1 \right) = \left( {{a_0} + ... + {{\left( {1 - {q_k}} \right)}^{{X_{\max }}}}{a_{{X_{\max }}}}} \right){P_{k - 1}}\left( 1 \right) + \\
\left( {\left( {\begin{array}{*{20}{c}}
1\\
1
\end{array}} \right){q_k}{a_1} + ...{\rm{ + }}\left( {\begin{array}{*{20}{c}}
{{X_{\max }}}\\
1
\end{array}} \right){q_k}{{\left( {1 - {q_k}} \right)}^{{X_{\max }} - 1}} \times } \right.\\
\left. {{a_{{X_{\max }}}}} \right){P_{k - 1}}\left( 0 \right),
\end{array}
\end{equation}
$\vdots $
\begin{equation}\label{eq:10}
\begin{array}{l}
{P_k}\left( {\left( {k - 1} \right){X_{\max }}} \right) = \\
\left( {{a_0} + ... + {{\left( {1 - {q_k}} \right)}^{{X_{\max }}}}{a_{{X_{\max }}}}} \right){P_{k - 1}}\left( {\left( {k - 1} \right){X_{\max }}} \right) + \\
\left( {\left( {\begin{array}{*{20}{c}}
1\\
1
\end{array}} \right){q_k}{a_1} + ...{\rm{ + }}\left( {\begin{array}{*{20}{c}}
{{X_{\max }}}\\
1
\end{array}} \right){q_k}{{\left( {1 - {q_k}} \right)}^{{X_{\max }} - 1}}} \right. \times \\
\left. {{a_{{X_{\max }}}}} \right){P_{k - 1}}\left( {\left( {k - 1} \right){X_{\max }} - 1} \right) + \\
\left( {\left( {\begin{array}{*{20}{c}}
2\\
2
\end{array}} \right)q_k^2{a_2} + ...{\rm{ + }}\left( {\begin{array}{*{20}{c}}
{{X_{\max }}}\\
2
\end{array}} \right)q_k^2{{\left( {1 - {q_k}} \right)}^{M - 3}}} \right. \times \\
\left. {{a_{{X_{\max }}}}} \right){P_{k - 1}}\left( {\left( {k - 1} \right){X_{\max }} - 2} \right) + \\
{\rm{                          }} \vdots \\
\left( {q_k^{{X_{\max }} - 1}{a_{{X_{\max }} - 1}} + \left( {\begin{array}{*{20}{c}}
{{X_{\max }}}\\
{{X_{\max }} - 1}
\end{array}} \right)q_k^{{X_{\max }} - 1}\left( {1 - {q_k}} \right)} \right. \times \\
\left. {{a_{{X_{\max }}}}} \right){P_{k - 1}}\left( {\left( {k - 1} \right){X_{\max }} - \left( {{X_{\max }} - 1} \right)} \right) + \\
q_k^{{X_{\max }}}{a_{{X_{\max }}}}{P_{k - 1}}\left( {\left( {k - 1} \right){X_{\max }} - {X_{\max }}} \right),
\end{array}
\end{equation}
\begin{eqnarray}\label{eq:11}
{P_k}\left( {\left( {k - 1} \right){X_{\max }} + 1} \right) = 0,
\end{eqnarray}

\section{Upper and Lower Bounds}
We now provide two lower and an upper bounds for MC in presence of ISI. While the expressions in~\eqref{eq:6} and~\eqref{eq:7} are valid for any $k$, for tractability, we now assume that the ISI only affects the next time-slot, i.e., molecules are received within two time slots or disappear. This is equivalent to $k = 2$ and we have $P_1(0) = 1$, ${P_1}\left( {n \ne 0} \right) = 0$.
\subsection{Lower bound 1}	
In the first lower bound we consider the effect of ISI degradation on mutual information between input and output symbols. This is a lower bound on the channel capacity because this measure ignores the memory; essentially, we consider a DMC but with an additional source of measurement error due to molecules from the previous timeslot. Indeed, comparing the capacity of the DMC ($C_{DMC}$) and this lower bound measures the error of ignoring ISI. This lower bound relates to lower bound of $I_{i.i.d}$ in discrete Gaussian channel with ISI in~\cite{10}. Hence we have
\begin{equation}\label{eq:12}
\begin{array}{l}
{I_{L{B_1}}} = I\left( {{X_m};{Y_m}} \right) = H\left( {{Y_m}} \right) - H\left( {\left. {{Y_m}} \right|{X_m}} \right)\\
{\rm{        = }} - \sum\limits_{{y_m} = 0}^{2{X_{\max }}} {p\left( {{y_m}} \right)\log \left( {p\left( {{y_m}} \right)} \right)}  + \\
{\rm{             }}\sum\limits_{{x_m} = 0}^{{X_{\max }}} {{a_{{x_m}}}} \sum\limits_{{y_m} = 0}^{{x_m} + {X_{\max }}} p \left( {\left. {{y_m}} \right|{x_m}} \right)\log p\left( {\left. {{y_m}} \right|{x_m}} \right),
\end{array}
\end{equation}
where $p\left( {\left. {{y_m}} \right|{x_m}} \right)$ is given by
\begin{equation}\label{eq:13}
p\left( {\left. {{y_m}} \right|{x_m}} \right) = \left\{ {\begin{array}{*{20}{c}}
\begin{array}{l}
\sum\limits_{i = 0}^{{x_m}} {\left( {\begin{array}{*{20}{c}}
{{x_m}}\\
i
\end{array}} \right){{\left( {1 - {q_1}} \right)}^{{x_m} - i}}q_1^i \times } \\
\sum\limits_{j = {y_m} - i}^{{X_{\max }}} {\left( {\begin{array}{*{20}{c}}
j\\
{{y_m} - i}
\end{array}} \right){a_j}} q_2^{{y_m} - i} \times \\
{\left( {1 - {q_2}} \right)^{j - \left( {{y_m} - i} \right)}}{\rm{            }},{y_m} \le {x_m} + {X_{\max }}
\end{array}\\
{0{\rm{                              ,}}{y_m} > {x_m} + {X_{\max }}.}
\end{array}} \right.
\end{equation}
By averaging over ${x_m}$ on $p\left( {\left. {{y_m}} \right|{x_m}} \right)$  , $p\left( {{y_m}} \right)$ is given by
\begin{equation}
\begin{array}{l}
p\left( {{y_m}} \right) = \sum\limits_{{x_m} = 0}^{{X_{\max }}} {{a_{{x_m}}}\sum\limits_{i = 0}^{{x_m}} {\left( {\begin{array}{*{20}{c}}
{{x_m}}\\
i
\end{array}} \right){{\left( {1 - q_1} \right)}^{{x_m} - i}}{q_1^i}}  \times } \\
{\rm{            }}\sum\limits_{j = y - i}^{{X_{\max }}} {\left( {\begin{array}{*{20}{c}}
j\\
{{y_m} - i}
\end{array}} \right){a_j}} q_2^{{y_m} - i}{\left( {1 - {q_2}} \right)^{j - \left( {{y_m} - i} \right)}}.
\end{array}
\end{equation}
\subsection{Lower Bound 2}
We assume that molecules are received within two time-slots or disappear; hence the transmitted symbol in current time-slot only affects received molecules in current and next time-slot. We consider mutual information between transmitted symbol in current time-slot and received symbols in current and next time-slot
\begin{equation}\label{eq:14}
\begin{array}{l}
{I_{L{B_2}}} = I\left( {{X_{m - 1}};{Y_{m - 1}},{Y_m}} \right) \\ \mathop  = \limits^{\left( {\rm{a}} \right)} H\left( {{Y_m},{Y_{m - 1}}} \right)
 - H\left( {\left. {{Y_m},{Y_{m - 1}}} \right|{X_{m - 1}}} \right) \\ \mathop  = \limits^{\left( {\rm{b}} \right)} H\left( {{Y_{m - 1}}} \right) + H\left( {\left. {{Y_m}} \right|{Y_{m - 1}}} \right) -
H\left( {\left. {{Y_{m - 1}}} \right|{X_{m - 1}}} \right) \\ \hspace*{0.3in} - H\left( {\left. {{Y_m}} \right|{X_{m - 1}},{Y_{m - 1}}} \right),
\end{array}
\end{equation}
where (a) is obtained based on definition of mutual information and (b) based on definition of joint entropy. We consider the channel in steady state regime, hence $P\left( {\left. {{y_{m - 1}}} \right|{x_{m - 1}}} \right) = p\left( {\left. {{y_m}} \right|{x_m}} \right)$, which is given in \eqref{eq:13}. Also, $p\left( {\left. {{y_m}} \right|{y_{m - 1}},{x_{m - 1}}} \right)$ is given by
\begin{equation}\label{eq:15}
\begin{array}{l}
p\left( {\left. {{y_m}} \right|{y_{m - 1}},{x_{m - 1}}} \right)\mathop  = \limits^{\left( {\rm{a}} \right)} \\
\sum\limits_{{x_m} = 0}^{{X_{\max }}} {p\left( {\left. {{x_m}} \right|{y_{m - 1}},{x_{m - 1}}} \right)p\left( {\left. {{y_m}} \right|{y_{m - 1}},{x_m},{x_{m - 1}}} \right)} \\ \mathop  = \limits^{\left( b \right)}
\sum\limits_{{x_m} = 0}^{{X_{\max }}} {p\left( {{x_m}} \right)p\left( {\left. {{y_m}} \right|{y_{m - 1}},{x_m},{x_{m - 1}}} \right),}
\end{array}
\end{equation}
where (a)  is obtained based on law of total probability, (b) is obtained based on independence of $x_m$ from $y_{m-1}$ and $x_{m-1}$ which is due to the i.i.d assumption of the input distribution and causality. By averaging over $x_{m-1}$ in \eqref{eq:15}, $p\left( {\left. {{y_m}} \right|{y_{m - 1}}} \right)$ is given by
\begin{equation}\label{eq:16}
\begin{array}{l}
p\left( {\left. {{y_m}} \right|{y_{m - 1}}} \right) = \\
\sum\limits_{{x_{m - 1}} = 0}^{{X_{\max }}} {p\left( {\left. {{x_{m - 1}}} \right|{y_{m - 1}}} \right)p\left( {\left. {{y_m}} \right|{y_{m - 1}},{x_{m - 1}}} \right)} \\
\mathop  = \limits^{\left( {\rm{a}} \right)} \sum\limits_{{x_{m - 1}} = 0}^{{X_{\max }}} {p\left( {\left. {{x_{m - 1}}} \right|{y_{m - 1}}} \right)\sum\limits_{{x_m} = 0}^{{X_{\max }}} {p\left( {{x_m}} \right)} } {\rm{ }} \times \\
p\left( {\left. {{y_m}} \right|{y_{m - 1}},{x_m},{x_{m - 1}}} \right)\mathop  = \limits^{\left( {\rm{b}} \right)} \sum\limits_{{x_{m - 1}} = 0}^{{X_{\max }}} {\frac{{p\left( {\left. {{y_{m - 1}}} \right|{x_{m - 1}}} \right)p\left( {{x_{m - 1}}} \right)}}{{p\left( {{y_{m - 1}}} \right)}} \times } \\
\sum\limits_{{x_m} = 0}^{{X_{\max }}} {p\left( {{x_m}} \right)p\left( {\left. {{y_m}} \right|{y_{m - 1}},{x_m},{x_{m - 1}}} \right)} .
\end{array}
\end{equation}
where (a) and (b) are obtained based on law of total probability, and Bayes'  rule, respectively. Also, $p\left( {\left. {{y_m}} \right|{y_{m - 1}},{x_m},{x_{m - 1}}} \right)$ is given by
\begin{equation}\label{eq:17}
\begin{array}{l}
p\left( {\left. {{y_m}} \right|{y_{m - 1}},{x_m},{x_{m - 1}}} \right)\\
\mathop  = \limits^{\left( {\rm{a}} \right)} \sum\limits_{{{y'}_{m - 2}} = 0}^{{X_{\max }}} {p\left( {\left. {{{y'}_{m - 2}}} \right|{y_{m - 1}},{x_m},{x_{m - 1}}} \right)}  \times \\
{\rm{          }}p\left( {\left. {{y_m}} \right|{y_{m - 1}},{x_m},{x_{m - 1}},{{y'}_{m - 2}}} \right)\\
\mathop  = \limits^{\left( {\rm{b}} \right)} \sum\limits_{{{y'}_{m - 2}} = 0}^{{X_{\max }}} {p\left( {\left. {{{y'}_{m - 2}}} \right|{y_{m - 1}},{x_{m - 1}}} \right)}  \times \\
{\rm{          }}p\left( {\left. {{y_m}} \right|{y_{m - 1}},{x_m},{x_{m - 1}},{{y'}_{m - 2}}} \right)\\
\mathop  = \limits^{\left( {\rm{c}} \right)} \sum\limits_{{{y'}_{m - 2}} = 0}^{{X_{\max }}} {\frac{{p\left( {\left. {{y_{m - 1}},{x_{m - 1}}} \right|{{y'}_{m - 2}}} \right)p\left( {{{y'}_{m - 2}}} \right)}}{{p\left( {{y_{m - 1}},{x_{m - 1}}} \right)}}}  \times \\
{\rm{          }}p\left( {\left. {{y_m}} \right|{y_{m - 1}},{x_m},{x_{m - 1}},{{y'}_{m - 2}}} \right)\\
\mathop  = \limits^{\left( {\rm{d}} \right)} \sum\limits_{{{y'}_{m - 2}} = 0}^{{X_{\max }}} {\frac{{p\left( {\left. {{y_{m - 1}}} \right|{{y'}_{m - 2}},{x_{m - 1}}} \right)p\left( {\left. {{x_{m - 1}}} \right|{{y'}_{m - 2}}} \right)p\left( {{{y'}_{m - 2}}} \right)}}{{p\left( {\left. {{y_{m - 1}}} \right|{x_{m - 1}}} \right)p\left( {{x_{m - 1}}} \right)}} \times } \\
{\rm{          }}p\left( {\left. {{y_m}} \right|{y_{m - 1}},{x_m},{x_{m - 1}},{{y'}_{m - 2}}} \right)\\
\mathop  = \limits^{\left( {\rm{e}} \right)} \sum\limits_{{{y'}_{m - 2}} = 0}^{{X_{\max }}} {\frac{{p\left( {\left. {{y_{m - 1}}} \right|{{y'}_{m - 2}},{x_{m - 1}}} \right)p\left( {{x_{m - 1}}} \right)p\left( {{{y'}_{m - 2}}} \right)}}{{p\left( {\left. {{y_{m - 1}}} \right|{x_{m - 1}}} \right)p\left( {{x_{m - 1}}} \right)}} \times } \\
{\rm{         }}p\left( {\left. {{y_m}} \right|{y_{m - 1}},{x_m},{x_{m - 1}},{{y'}_{m - 2}}} \right)\\
 = \sum\limits_{{{y'}_{m - 2}} = 0}^{{X_{\max }}} {\frac{{p\left( {\left. {{y_{m - 1}}} \right|{{y'}_{m - 2}},{x_{m - 1}}} \right)p\left( {{{y'}_{m - 2}}} \right)}}{{p\left( {\left. {{y_{m - 1}}} \right|{x_{m - 1}}} \right)}} \times } \\
{\rm{           }}p\left( {\left. {{y_m}} \right|{y_{m - 1}},{x_m},{x_{m - 1}},{{y'}_{m - 2}}} \right)\\
\mathop  = \limits^{\left( {\rm{f}} \right)} \sum\limits_{{{y'}_{m - 2}} = 0}^{{y_{m - 1}}} {\frac{{p\left( {\left. {{{x'}_{m - 1}}} \right|{x_{m - 1}}} \right)p\left( {{{y'}_{m - 2}}} \right)}}{{p\left( {\left. {{y_{m - 1}}} \right|{x_{m - 1}}} \right)}}}  \times {\rm{  }}\\
{\rm{            }}p\left( {\left. {{y_m}} \right|{y_{m - 1}},{x_m},{x_{m - 1}},{{y'}_{m - 2}}} \right),{\rm{   }}
\end{array}
\end{equation}
where (a) is obtained based on law of total probability and ${y'_{m - 2}}$ is the number of received molecules at the end of time-slot $m-1$ from transmitted molecules in time-slot $m-2$; (b) is obtained due to the independence of ${y'_{m - 2}}$ from $x_m$; (c) is obtained based on Bayes' rule; (d) is obtained based on the joint probability formula, (e) is obtained based on independence of ${y'_{m - 2}}$ and $x_{m-1}$  and the joint probability formula; (f) is obtained due to the fact that ${y_{m - 1}} = {x'_{m - 1}} + {y'_{m - 2}}$, where ${x'_{m - 1}}$ denotes the number of absorbed molecules at time-slot $m-1$ at end of time-slot $m-1$. Based on definition of ${y'_{m - 2}}$ we have
\begin{equation}\label{eq:18}
\begin{array}{l}
p\left( {{{y'}_{m - 2}}} \right) = \sum\limits_{{x_{m - 2}} = 0}^{{X_{\max }}} {{a_{{x_{m - 2}}}}\left( {\begin{array}{*{20}{c}}
{{x_{m - 2}}}\\
{{{y'}_{m - 2}}}
\end{array}} \right)} {q_2}^{{{y'}_{m - 2}}} \times \\
{\left( {1 - {q_2}} \right)^{{x_{m - 2}} - {{y'}_{m - 2}}}},
\end{array}
\end{equation}
\begin{equation}\label{eq:19}
\begin{array}{l}
p\left( {\left. {{y_m}} \right|{y_{m - 1}},{x_m},{x_{m - 1}},{{y'}_{m - 2}}} \right)\mathop  = \limits^{\left( {\rm{a}} \right)} \\
p\left( {\left. {{y_m}} \right|{x_m},{{x''}_{m - 1}} = {x_{m-1}} - \left( {{y_{m - 1}} - {{y'}_{m - 2}}} \right)} \right) = \\
\left\{ {\begin{array}{*{20}{c}}
\begin{array}{l}
\sum\limits_{i = {y_m} - {x_{m - 1}}}^{{x_m}} {\left( {\begin{array}{*{20}{c}}
{{x_m}}\\
i
\end{array}} \right){{\left( {1 - {q_1}} \right)}^{{x_m} - i}}{q_1}^i\left( {\begin{array}{*{20}{c}}
{{{x''}_{m - 1}}}\\
{{y_m} - i}
\end{array}} \right) \times } \\
q_2^{{y_m} - i}{\left( {1 - {q_2}} \right)^{{{x''}_{m - 1}} - \left( {{y_m} - i} \right)}}{\rm{,}}{y_m} < {x_m} + {{x''}_{m - 1}}
\end{array}\\
{{\rm{ }}\begin{array}{*{20}{c}}
0&{}&{}&{}&{}&{}
\end{array}{\rm{,}}{y_m} > {x_m} + {{x''}_{m - 1}}},
\end{array}} \right.
\end{array}
\end{equation}
(a) is obtained because ${x_{m - 1}} = {x'_{m - 1}} + {x''_{m - 1}}$, where ${x''_{m - 1}}$ is the number of remaining molecules at end of time-slot $m-1$   from transmitted molecules in the same time-slot. Moreover, based on definition of ${x'_{m - 1}}$ , $P\left( {\left. {{{x'}_{m - 1}}} \right|{x_{m - 1}}} \right)$ is given by
\begin{equation}\label{eq:20}
P\left( {\left. {{{x'}_{m - 1}}} \right|{x_{m - 1}}} \right) = \left( {\begin{array}{*{20}{c}}
{{x_{m - 1}}}\\
{{{x'}_{m - 1}}}
\end{array}} \right)q_1^{{{x'}_{m - 1}}}{\left( {1 - {q_1}} \right)^{{x_{m - 1}} - {{x'}_{m - 1}}}}.
\end{equation}
\subsection{Upper Bound}
We assume that molecules are received or disappear after two time-slots. Hence, if we send symbols and wait two time-slots before transmitting the next symbol we have an interference-free channel. This is equivalent to the DMC case with the binomial transition probabilities of~\eqref{eq:2} where $q_1$ is replaced by $q_U = F_W(2T)$. The mutual information of this channel is concave which can be maximized using the Blahut-Arimuto algorithm~(BAA) \cite{17}. We therefore have an upper bound as follows
\begin{equation}\label{eq:21}
{I_{UB}} = \max_{\textbf{a}} I\left( {{X_m};{Y_m}} \right) = \max_{\textbf{a}} \left[H\left( {{Y_m}} \right) - H\left( {\left. {{Y_m}} \right|{X_m}} \right)\right],
\end{equation}
where $P\left( {\left. {{y_m}} \right|{x_m}} \right)$ is given by
\begin{equation}\label{eq:22}
P\left( {\left. {{y_m}} \right|{x_m}} \right) = \left\{ {\begin{array}{*{20}{c}}
{\left( {\begin{array}{*{20}{c}}
{{x_m}}\\
{{y_m}}
\end{array}} \right)q_U^{^{{y_m}}}{{\left( {1 - {q_U}} \right)}^{{x_m} - {y_m}}}{\rm{,}}{y_m} \le {x_m}}\\
{\begin{array}{*{20}{c}}
0&{}&{}&{}&{}&{}&{}
\end{array}{\rm{ ,}}{y_m} > {x_m}.}
\end{array}} \right.
\end{equation}
and ${q_U} = {F_W}\left( {2T} \right)$.
This upper bound relates to "matched filter bound" of discrete Gaussian channel with ISI in \cite{10}.
\subsection{Matched Filter}
We now provide the achievable rate when using a matched filter. We transmit symbols and wait two time-slots before transmitting another symbol. We again have an interference-free channel. Naturally, although this scheme may be appealing in a bit per transmission (channel use) sense, it pays a toll when we consider the rate (in bits per second) as it uses two time slots. By this definition, ${I_{MF}}$ is given by
\begin{equation}\label{eq:23}
\begin{array}{l}
{I_{MF}} = I\left( {{X_{m - 1}};{Y_{m - 1}},{Y_m}} \right)\mathop  = \limits^{\left( {\rm{a}} \right)} H\left( {{Y_m},{Y_{m - 1}}} \right) - \\
H\left( {\left. {{Y_m},{Y_{m - 1}}} \right|{X_{m - 1}}} \right)\mathop  = \limits^{\left( {\rm{b}} \right)} H\left( {{Y_{m - 1}}} \right) + H\left( {\left. {{Y_m}} \right|{Y_{m - 1}}} \right) - \\
H\left( {\left. {{Y_{m - 1}}} \right|{X_{m - 1}}} \right) - H\left( {\left. {{Y_m}} \right|{X_{m - 1}},{Y_{m - 1}}} \right),
\end{array}
\end{equation}
where (a) is obtained from the definition of mutual information and (b) from the joint entropy formula. We measure the number of molecules at end of each time-slot. In this case $P\left( {\left. {{y_{m - 1}}} \right|{x_{m - 1}}} \right)$ is given by
\begin{equation}\label{eq:24}
\begin{array}{l}
P\left( {\left. {{y_{m - 1}}} \right|{x_{m - 1}}} \right) = \\
\left\{ {\begin{array}{*{20}{c}}
{\left( {\begin{array}{*{20}{c}}
{{x_{m - 1}}}\\
{{y_{m - 1}}}
\end{array}} \right)q_1^{{y_{m - 1}}}{{\left( {1 - {q_1}} \right)}^{{x_{m - 1}} - {y_{m - 1}}}},{y_{m - 1}} \le {x_{m - 1}},}\\
{\begin{array}{*{20}{c}}
0&{}&{}&{}&{}&{}&{}&{}
\end{array}{\rm{,}}{y_{m - 1}} > {x_{m - 1}}.}
\end{array}} \right.
\end{array}
\end{equation}
Also, $p\left( {\left. {{y_m}} \right|{y_{m - 1}},{x_{m - 1}}} \right)$ for ${y_{m - 1}} < {x_{m - 1}}$ is given by
\begin{equation}\label{eq:25}
\begin{array}{l}
p\left( {\left. {{y_m}} \right|{y_{m - 1}},{x_{m - 1}}} \right) = \\
\left\{ {\begin{array}{*{20}{c}}
\begin{array}{l}
\left( {\begin{array}{*{20}{c}}
{{x_{m - 1}} - {y_{m - 1}}}\\
{{y_m}}
\end{array}} \right){q_2}^{{y_m}} \times \\
{\left( {1 - {q_2}} \right)^{{x_{m - 1}} - {y_{m - 1}} - {y_m}}}{\rm{, }}{y_m} \le {x_{m - 1}} - {y_{m - 1}}
\end{array}\\
{\begin{array}{*{20}{c}}
0&{}&{}&{}&{}
\end{array}{\rm{, }}{y_m} > {x_{m - 1}} - {y_{m - 1}},}
\end{array}} \right.
\end{array}
\end{equation}
Using law of total probability on \eqref{eq:25} we have
\begin{equation}\label{eq:26}
\begin{array}{l}
p\left( {\left. {{y_m}} \right|{y_{m - 1}}} \right) = \sum\limits_{{x_{m - 1}} = 0}^{{X_{\max }}} {p\left( {\left. {{y_m}} \right|{y_{m - 1}},{x_{m - 1}}} \right)p\left( {\left. {{x_{m - 1}}} \right|{y_{m - 1}}} \right)} \\
\mathop  = \limits^{\left( {\rm{a}} \right)} \sum\limits_{{x_{m - 1}} = 0}^{{X_{\max }}} {p\left( {\left. {{y_m}} \right|{y_{m - 1}},{x_{m - 1}}} \right)\frac{{p\left( {\left. {{y_{m - 1}}} \right|{x_{m - 1}}} \right)p\left( {{x_{m - 1}}} \right)}}{{p\left( {{y_{m - 1}}} \right)}}},
\end{array}
\end{equation}
where (a) is obtained based on Bayes' rule, also, $I_{MF}$ is calculated by derived $\textbf{a}$ from optimizing $C_{DMC}$. Clearly,
\begin{equation}\label{eq:27}
\begin{array}{l}
{I_{L{B_2}}} = I\left( {{X_{m - 1}};{Y_{m - 1}},{Y_m}} \right)\mathop  = \limits^{\left( {\rm{a}} \right)} I\left( {{X_{m - 1}};{Y_{m - 1}}} \right) + \\
I\left( {{X_{m - 1}};{Y_m}\left| {{Y_{m - 1}}} \right.} \right)\mathop  \ge \limits^{\left( b \right)} I\left( {{X_{m - 1}};{Y_{m - 1}}} \right) = {I_{L{B_1}}},
\end{array}
\end{equation}
where (a) is obtained from chain rule in mutual information and (b) is obtained based on non-negativity assumption of mutual information. Hence, we have following result
\[{I_{L{B_1}}} \le {I_{L{B_2}}} \le {I_{i.i.d}} \le {I_{UB}}. \]
\subsection{Optimizing The Lower and Upper Bounds}
We are not able to show the concavity of the lower bounds ${I_{L{B_1}}}$  and ${I_{L{B_2}}}$  with respect to $a_{x_m}$; however, we can modify the BAA \cite{17} to find a local maximum, and as a result, we can calculate these two lower bounds. Each element of channel transition probablity matrices ${{\bf{P}}^{\left( {L{B_h}} \right)}}$, $h \in \left\{ {1,2} \right\}$ for lower bounds 1 and 2 are given by
\begin{equation}\label{eq:29}
{{\bf{P}}_{{x_m},x{y_m}}^{\left( {{LB}_1} \right)}} = \left[ {p\left( {\left. {{Y_m} = {y_m}} \right|{X_{m}} = {x_m}} \right)} \right],
\end{equation}
\begin{equation}\label{eq:30}
{{\bf{P}}_{{x_{m-1}},{y_{m-1}y_m}}^{\left( {{LB}_2} \right)}} =  \left[ {p\left( {\left. {{Y_m} = {y_m},{Y_{m - 1}} = {y_{m - 1}}} \right|{X_{m - 1}} = {x_{m - 1}}} \right)} \right]
\end{equation}
The size of ${{\bf{P}}^{\left( {{LB}_1} \right)}}$  and ${{\bf{P}}^{\left( {{LB}_2} \right)}}$  are  $\left( {{X_{\max }} + 1} \right) \times \left( {2{X_{\max }} + 1} \right)$ and  $\left( {{X_{\max }} + 1} \right) \times {\left( {2{X_{\max }} + 1} \right)^2}$ respectively. Also, ${y_{m,}}{y_{m - 1}} \in \left[ {0,...,2{X_{\max }}} \right]$   and ${x_m} \in \left[ {0,...,{X_{\max }}} \right]$ . For matrix ${\bf{Q}}$ with size of ${{\bf{P}}^{\left( {{LB}_h} \right)}}$, $h \in \left\{ {1,2} \right\}$ , let
\begin{equation}\label{eq:31}
J\left( {{\bf{a}},{{\bf{P}}^{\left({LB}_h\right)}},{\bf{Q}}} \right) = \sum\nolimits_j {\sum\nolimits_i {{a_j}{\bf{P}}_{j,i}^{\left({LB}_h\right)}\log \frac{{{{\bf{Q}}_{i,j}}}}{{{a_j}}}} } .
\end{equation}
 where $\textbf{Q}_{i,j}$ denotes the element $(i,j)^{\emph{th}}$ of $\textbf{Q}$. Then the following is true
 \begin{enumerate}
   \item $I_{{LB}_{h}} = \mathop {\max }\limits_{\bf{a}} \mathop {\max }\limits_{\bf{Q}} J\left( {{\bf{a}},{\bf{P}}_{}^{\left({LB}_h\right)},{\bf{Q}}} \right).$
   \item For fixed ${\bf{a}}$, $J\left( {{\bf{a}},{\bf{P}}^{\left({LB}_h\right)},{\bf{Q}}} \right)$ is locally maximized by
   \begin{equation}\label{eq:32}
   {{\bf{Q}}_{i,j}} = \frac{{{a_j}{\bf{P}}_{j,i}^{\left({LB}_h\right)}}}{{\sum\nolimits_j {{a_j}{\bf{P}}_{j,i}^{\left({LB}_{h}\right)}} }}.
   \end{equation}
   \item For fixed ${\bf{Q}}$, $J\left( {{\bf{a}},{\bf{P}}_{}^{\left({LB}_h\right)},{\bf{Q}}} \right)$  is locally maximized by
   \begin{equation}\label{eq:33}
   {a_j} = \frac{{\exp \left( {\sum\nolimits_i {{\bf{P}}_{j,i}^{\left({LB}_{h}\right)}\log {{\bf{Q}}_{i,j}}} } \right)}}{{\sum\nolimits_j {\exp \left( {\sum\nolimits_i {{\bf{P}}_{j,i}^{\left({LB}_h\right)}\log {{\bf{Q}}_{i,j}}} } \right)} }}
   \end{equation}
 \end{enumerate}
The algorithm iterates between $a_j$ derived in \eqref{eq:33} and the transition probability matrices ${{\bf{P}}^{\left( {L{B_h}} \right)}}$ in \eqref{eq:29} or \eqref{eq:30}. This procedure is repeated until the convergence of ${a_j}$.

In contrast to the lower bounds, ${I_{UB}}$ is a concave function in terms of ${a_j}$. Hence using the standard BAA~\cite{17} the upper bound can be maximized.
\section{Numerical Results}
In this Section the bounds of information rate of ASK modulation in channel with ISI and AIGN model for transmission time are evaluated and compared with each other, $I_{MF}$ and the $C_{DMC}$. Note that in the DMC case, molecules arrive within their time-slot or disappear; as a result the corresponding performance curves are simply provided for insight and context on those of the ISI channel, as opposed to a direct comparison. We also study the effect of parameters in molecular medium such as $l$, $v$ and $\sigma^2$.

Fig.~1 plots ${I_{{LB}_2}}$ and ${I_{{LB}_1}}$  as a function of $T$ with $v = 1$, $\l = 10^{-2}$ and $\sigma^2 = 1$ for the optimized input distribution from~\eqref{eq:33} and a uniform distribution. It can seen that using the optimized distribution, the bounds ${I_{L{B_2}}}$ and ${I_{L{B_1}}}$ is improved significantly. Moreover using the optimized distribution of~\eqref{eq:33} increases the difference between ${I_{{LB}_2}}$ and ${I_{{LB}_1}}$ to increase in comparison with the uniform input.

Fig.~2 plots ${I_{{LB}_2}}$, ${I_{{LB}_1}}$, ${I_{MF}}$, ${C_{DMC}}$ and ${I_{UB}}$ versus $T$ for different values of $l$ with $v = 1$, $\sigma^2 = 1$ and $X_{\max} = 7$. It can be observed that by decreasing $l$ all bounds increase and converge to ${\log _2}\left( {{X_{\max }} + 1} \right) = 3$ , which is the entropy of the source. Crucially, over wide ranges the upper and lower bounds are close. Also, due to reduced effect of ISI, by reducing $l$ the gap between all derived bounds is reduced. Furthermore, Since by increasing $T$ the probability of receiving molecules in current time slot is increased, $I_{MF}$ also converges to $C_{DMC}$ for larger values of $T$.

Fig.~3 plots  ${I_{{LB}_2}}$, ${I_{{LB}_1}}$, ${I_{MF}}$, ${C_{DMC}}$ and ${I_{UB}}$ versus $T$ for different values of $\sigma^2$ with $v = 1$ , $l = 10^{-2}$ and $X_{\max} = 7$. All capacity bounds are increasing functions of $\sigma$. While this seems counter intuitive, an increasing $\sigma$ causes reduced ISI - this is consistent with this is consistent with the fact that $q_1$ is an increasing function of $\sigma$ for low-to-medium values of $v$, i.e., the increased variance in position increases the probability of the molecules arriving with two time-slots after release. For large $v$, $q_1 \simeq 1$ and ISI is not really an issue. It is worth noting that when information encoded in time-of-release as in~\cite{2}, the mutual information is not monotonic in $\sigma$.

Fig.~4 plots ${I_{{LB}_2}}$, ${I_{{LB}_1}}$, ${I_{MF}}$, ${C_{DMC}}$ and ${I_{UB}}$ versus $T$ for different values of $v$ with $\sigma^2 = 1$ , $l = 10^{-2}$ and $X_{\max} = 7$ . As in~\cite{2}, increasing drift velocity increases mutual information due to reduced ISI.

Comparison the results of the three recent figures, we note that the bounds are most sensitive to the transmitter-receiver distance $l$ and drift velocity $v$ while not being as sensitive to the diffusion constant $\sigma$. Our numerical and simulation studies based on performance of an ML (maximum likelihood) detector in different settings (not reported here), suggests that indeed the presented ISI model with one time-slot memory is viable over a wide range of time slot durations, $T$.

 \begin{figure}
 \centerline{
    \epsfxsize = 3.2in
    \epsffile{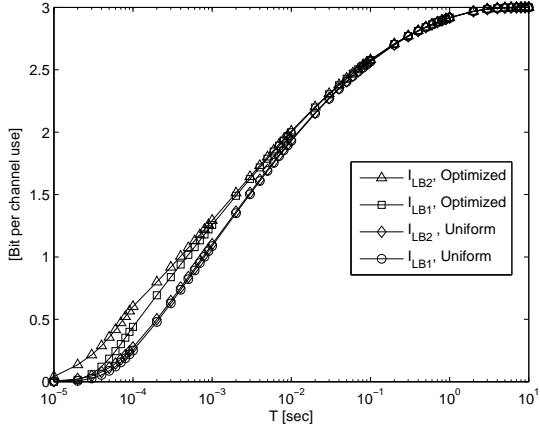}
    }
 \caption { ${I_{L{B_2}}}$ and ${I_{L{B_1}}}$ in terms of $T$ with $v = 1$, $l = 10^{-2}$, $\sigma^2 = 1$, $X_{max} = 7$ for optimized input distribution and uniform distribution.} \label{planBexhas_1}
 \end{figure}
 \begin{figure}
 \centerline{
    \epsfxsize = 3.2in
    \epsffile{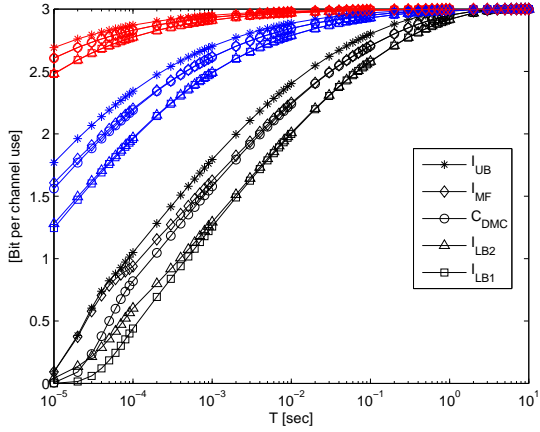}
    }
 \caption { ${I_{L{B_2}}}$, ${I_{L{B_1}}}$, ${I_{MF}}$, ${C_{DMC}}$ and ${I_{UB}}$ in terms of $T$ for different values of $l$ (black curves, $l = 10^{-2}$, blue curves, $l = 10^{-3}$, red curves, $l = 10^{-4}$) with $v = 1$, $\sigma^2 = 1$, $X_{max} = 7$.} \label{planBexhas_2}
 \end{figure}
 \begin{figure}
 \centerline{
    \epsfxsize = 3.2in
    \epsffile{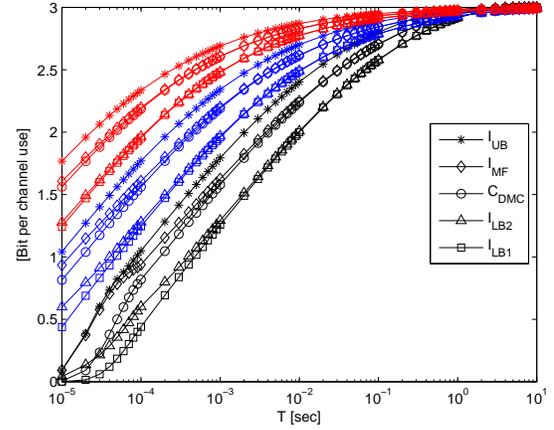}
    }
 \caption { ${I_{L{B_2}}}$, ${I_{L{B_1}}}$, ${I_{MF}}$, ${C_{DMC}}$ and ${I_{UB}}$ in terms of $T$ for different values of $\sigma^2$ (black curves, $\sigma^2 = 1$, blue curves, $\sigma^2 = 10$, red curves, $\sigma^2 = 100$) with $v = 1$, $\l = 10^{-2}$, $X_{max} = 7$.} \label{planBexhas_3}
 \end{figure}
 \begin{figure}
 \centerline{
    \epsfxsize = 3.2in
    \epsffile{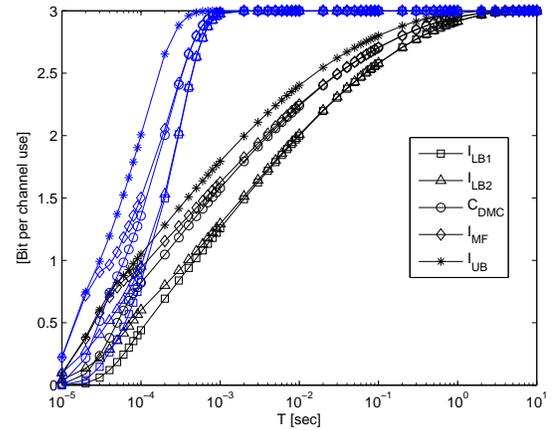}
    }
 \caption { ${I_{L{B_2}}}$, ${I_{L{B_1}}}$, ${I_{MF}}$, ${C_{DMC}}$ and ${I_{UB}}$ in terms of $T$ for different values of $v$ (black curves, $v = 1$, blue curves, $v = 100$) with $\l = 10^{-2}$, $\sigma^2 = 1$ and $X_{max} = 7$.} \label{planBexhas_4}
 \end{figure}
\section{Conclusions}
In this paper we consider ASK-based MC with ISI. Specifically, we derived two lower bounds and an upper bound on capacity (albeit with the significant simplification of i.i.d.~inputs). Our results showed the lower bounds are improved using an optimized distribution for the input probability of symbols. Also, our results showed that over wide parameter values the lower and upper bounds are close. The analysis presented here is useful in understanding the sensitivity of a MC system to the various parameters.

\bibliographystyle{IEEEtran}
\bibliography{IEEE_JSAC}

\end{document}